\documentstyle[amsmath,amssymb,epsf,mncite]{mn}

\pubyear{1999}
\volume{666}
\pagerange{\pageref{firstpage}--\pageref{lastpage}}
\begin{document}

\label{firstpage}

\title[Quantifying the evolution of higher--order clustering]{ 
Quantifying the evolution of higher--order clustering }

\newcounter{fusspilz}\setcounter{fusspilz}{0}
\def\fusspilz{{\stepcounter{fusspilz}\fnsymbol{fusspilz}}}
\author[Jens Schmalzing {\em et~al.}]{
Jens Schmalzing$^{1,2,\fusspilz}$,
Stefan Gottl\"ober$^{3,4,\fusspilz}$,
Anatoly A.\ Klypin$^{4,\fusspilz}$, and\cr
Andrey V.\ Kravtsov$^{4,\fusspilz}$
\setcounter{fusspilz}{0}
\vspace*{1ex}
\\
$^1$
Max--Planck--Institut f\"ur Astrophysik,
Karl--Schwarzschild--Stra{\ss}e 1,
85740 Garching, Germany.
\\
$^2$
Ludwig--Maximilians--Universit\"at,
Theresienstra{\ss}e 37,
80333 M\"unchen, Germany. 
\\
$^3$
Astrophysikalisches Institut Potsdam,
An der Sternwarte 16,
14482 Potsdam, Germany. 
\\
$^4$
Astronomy Department,
New Mexico State University, Dept.\ 4500,
Las Cruces, NM 88003, United States.
\\
$^\fusspilz$email jensen@mpa-garching.mpg.de \\
$^\fusspilz$email sgottloeber@aip.de \\
$^\fusspilz$email aklypin@nmsu.edu \\
$^\fusspilz$email akravtso@nmsu.edu
}

\date{Version of 22 June 1999.  Accepted for publication in Monthly Notices.}

\maketitle

% lengths
\parindent0ex
\parsep0ex
\parskip1ex
\voffset-.5in
\raggedbottom

% pictures
\def\bildchen#1{\begin{minipage}{\linewidth}\epsfxsize=\linewidth\epsfbox{#1}\end{minipage}}

% units of measurement
\def\hkpc{\ifmmode{h^{-1}{\rm kpc}}\else{$h^{-1}{\rm kpc}$}\fi}
\def\hMpc{\ifmmode{h^{-1}{\rm Mpc}}\else{$h^{-1}{\rm Mpc}$}\fi}
\def\hMsun{\ifmmode{h^{-1}M_\odot}\else{$h^{-1}M_\odot$}\fi}
\def\kms{{\rm km}/{\rm s}}

% mathematical macros
\def\ddx{{{\rm d}^d\!x\,}}

% letters in various styles
\def\bx{{\bf x}}
\def\cD{{\mathcal D}}
\def\rd{{\rm d}}
\def\re{{\rm e}}
\def\lV{{\overline V}}

\begin{abstract}
We  use a  high--resolution  dissipationless simulation  to study  the
evolution of  the dark  matter and halo  distributions in  a spatially
flat cosmological model dominated by a cosmological constant $\Lambda$
and  cold  dark  matter  ($\Lambda$CDM).   In order  to  quantify  the
evolution of structure, we  calculate the Minkowski functionals of the
halos  and  the  dark   matter  component  at  various  redshifts.   A
comparison of Minkowski functionals  and the more standard correlation
function  analysis  shows   that  the  Minkowski  functionals  contain
information  about correlation  functions of  arbitrary  order.  While
little evolution  of the Minkowski functionals of  halos between $z=4$
and $z=0$ is  observed, we find that the  Minkowski functionals of the
dark matter evolve rapidly with time.  The difference of the Minkowski
functionals of  halos and  dark matter can  be interpreted as  a scale
dependent bias.  This implies that scale--dependent bias is a property
of  not only  the two-point  halo  correlation function,  but also  of
correlation functions of higher order.
\end{abstract}

\begin{keywords}
{ Methods: statistical; Galaxies: general; Cosmology: theory; dark
matter; large--scale structure of Universe. }
\end{keywords}

\section{Introduction}

It is  generally believed  that dark matter  (DM) constitutes  a large
fraction  of the mass  in the  Universe.  Therefore,  it significantly
affects  both the process  of galaxy  formation, and  the large--scale
distribution of  galaxies.  The gravitational domination of  DM on the
scale  of  the  galactic  virial  radius  implies  that  collisionless
simulations can  be used  to study  the formation of  the DM  halos of
galaxies.

The standard  flat cosmological scenarios with cold  dark matter alone
cannot explain  the structure formation  both on small and  very large
scales.  Recently, scenarios with a non-zero vacuum energy, quantified
by the  cosmological constant $\Lambda$,  have been proved to  be very
successful in  describing most of  the observational data at  both low
and high  redshifts.  Here we study  the evolution of  the dark matter
and  halo  distributions  in   a  spatially  flat  cosmological  model
dominated  by   the  cosmological  constant  and   cold  dark  matter,
$\Lambda$CDM.

In  order  to  assess  the  properties  of the  dark  matter  or  halo
distribution, a  variety of  statistical methods have  been developed.
The hierarchy of correlation functions, counts--in--cells or the power
spectrum  are examples of  low--order statistics.   While this  may be
sufficient to characterize the large--scale structure of the Universe,
gravitational clustering quickly leads into the nonlinear regime where
strongly non--Gaussian  structures evolve.  More  information can then
be  expected  from  measures  that  provide a  handle  on  the  global
morphology  of  the  large--scale  structure.   One  example  of  such
measures are the Minkowski functionals.

Minkowski   functionals  have  been   introduced  into   cosmology  by
{\scite{mecke:robust}}.   They provide  a geometrical  and topological
description of a point  distribution (for a detailed introduction, see
{\pcite{schmalzing:minkowski}}).  Recently  Minkowski functionals have
been  applied to  clusters of  galaxies  {\cite{kerscher:abell}}, IRAS
galaxies  {\cite{kerscher:fluctuations}}, and  anisotropy maps  of the
cosmic                       microwave                      background
{\cite{winitzki:minkowski,schmalzing:minkowski_cmb,novikov:minkowski}}.

Our  article is  organized as  follows.  In  section~\ref{sec:simu} we
summarize  the properties  of  our $\Lambda$CDM  simulation, while  in
section~\ref{sec:mink}  we briefly introduce  the method  of Minkowski
functional analysis.   In the following  section~\ref{sec:analysis} we
apply  Minkowski  functionals to  describe  the  distribution of  dark
matter particles and halos.  Finally, in section~\ref{sec:conclusions}
we summarize obtained results and draw our conclusions.

\section{The simulation}
\label{sec:simu}

We  investigate a spatially  flat model  with a  cosmological constant
($\Lambda$CDM     with     $\Omega_0=0.3$,     $\Omega_{\Lambda}=0.7$,
$\sigma_8=1.0$,  and $h=0.7$).   These  parameters allow  for a  power
spectrum normalization  in accord with  both the four year  {\sl COBE}
DMR observations {\cite{bunn:fouryear}}  and the observed abundance of
galaxy clusters  {\cite{viana:cluster}}.  The  age of the  universe in
this model is $\approx 13.5$ Gyrs.

In order to study the statistical properties of halos in a
cosmological environment, the simulation box should be sufficiently
large; we use a cube of 60\hMpc\ side length.  On the other hand, to
assure halo survival in clusters, the force resolution should be
$\lesssim1-3\hkpc$ and the mass resolution should be
$\lesssim10^9\hMsun$ {\cite{moore:destruction,klypin:galaxies}}.  Our
simulations were done using the Adaptive Refinement Tree (ART) code
{\cite{kravtsov:adaptive}} with a dynamical range of 32,000 in high
density regions.  The simulation followed evolution of $256^3$ cold
dark matter particles, which give particle mass of
$1.1\times10^9\hMsun$.  The mass resolution is therefore sufficient to
identify galaxy--size halos of mass $M\gtrsim
5-10\times10^{10}\hMsun$.

Identification of  halos in  dense environments and  reconstruction of
their  evolution is  a challenge.   A variety  of situations  that are
frequently found  in the  real Universe are  difficult to  identify in
simulations.  Typically, problems arise  if a small satellite is bound
to a larger galaxy; famous examples are the M51 system, or the LMC and
the  Milky Way.   Cases  when many  small gravitationally  self--bound
halos are moving within a  large region of virial overdensity, such as
clusters  and groups  of galaxies,  can also  cause trouble.   We have
developed  two algorithms  to deal  with such  situations,  namely the
hierarchical friends-of-friends (HFOF) procedure and the bound density
maxima   (BDM)   algorithm   {\cite{klypin:galaxies}}.    Both   yield
consistent results, i.e.\ they basically  identify the same halos in a
DM simulation. 

In this study, we use the  BDM algorithm to construct halo catalogs at
different  redshifts.  The main  idea  of  the  algorithm is  to  find
positions of local  maxima in the density field  smoothed on a certain
scale and to  apply physically motivated criteria to  test whether the
identified  site  corresponds to  a  gravitationally  bound halo.  The
density maxima are found iteratively for a series of smoothing scales,
the  smallest  scale  being  close  to  the  peak  resolution  of  the
simulation. To  test whether a  density maximum corresponds to  a real
halo, we  construct density and velocity  dispersion profiles centered
on  the  maximum  and   iteratively  remove  unbound  particles.  This
procedure  eliminates  unbound  halos  from the  catalog,  while  also
cleaning bound halos off  some high-velocity background particles. The
detailed   description   of   the    BDM   algorithm   is   given   in
{\cite{klypin:galaxies}}  and  in  {\scite{colin:evolution}}. The  BDM
code is available from authors upon request.

Halos are usually characterized  by their virial mass.  Unfortunately,
it is difficult to obtain a meaningful definition of this quantity for
halos inside larger halos, so  we avoid the problem by determining the
maximum  circular  velocity   $v_{\text{circ}}$  of  each  halo.   The
circular  velocity is  defined  as the  maximum of  $\sqrt{GM(<R)/R}$,
where $M(<R)$ denotes the mass contained inside a sphere of radius $R$
around  the  halo  center,  and  $G$ is  the  gravitational  constant.
$v_{\text{circ}}$ is more useful both observationally and numerically.
Moreover,  it can  be measured  more easily  and more  accurately than
mass. For isolated halos the  maximum circular velocity and the virial
radius are, of course, directly related and are therefore equivalent.

The output  of the halo  finding algorithm is primarily  determined by
the  assumed lower  mass  threshold  and also  depends  weakly on  the
assumed maximum  halo radius.   With the threshold  of $10^{10}\hMsun$
(10 particles) at $z=0$ the algorithm identifies about 17,000 halos in
our simulation.   In order to conduct  a study of  halo statistics one
needs a  complete halo  sample that is  not affected by  the numerical
details  of   the  halo  finding   procedure.   We  have   tested  the
completeness  of  the  halo  samples with  the  differential  velocity
functions  {\cite{gottloeber:halo}}.  Depending  on  redshift one  can
define a  threshold of  the circular velocities  so that  for circular
velocities larger than  this threshold the halo samples  do not depend
on the numerical parameters of  the halo finder.  It has been verified
that   at  $z=0$   the  sample   is  complete   for  all   halos  with
$v_{\text{circ}}\gtrsim85\kms$,   while  for  $z=1$   completeness  is
achieved only with $v_{\text{circ}}\gtrsim100\kms$.

\section{Minkowski functionals}
\label{sec:mink}

Let us consider an  object in three--dimensional Euclidean space.  Its
morphology  can be  characterized by  the four  Minkowski functionals,
namely  the  volume $V$,  the  surface  area  $A$, the  integral  mean
curvature    $H$,    and     the    Euler    characteristic    $\chi$.
{\scite{hadwiger:vorlesung}} has shown  that the Minkowski functionals
supply a complete and unique description of the global morphology.  As
geometrical  characteristics of  structure, the  Minkowski functionals
combine both  the advantage of a  intuitive geometrical interpretation
and the advantage of delivering a quantitative measure.

In  the following, we  look upon  the distribution  of the  dark matter
particles  and  the halos  as  realizations  of  point processes.   To
investigate the distribution of these  points we decorate each of them
with a  ball of  radius $r$.  Starting  from the trivial  situation at
$r=0$, where all points are isolated from each other, we increase $r$,
thereby   creating  connections   between  neighboring   balls.   This
procedure  results in  a  complex pattern  of  intersecting balls  the
properties of which can be studied using Minkowski functionals.  To be
precise,  we  calculate for  a  set  of points  $\left\{\bx_j\right\}$
decorated  by  balls  $B$  of  radius $r$  the  Minkowski  functionals
$V_\mu\left(\bigcup_jB_{\bx_j}\right)$ of the  union set of all balls.
Varying  the  radius  introduces   a  diagnostic  parameter  into  our
investigation.  Since the  set of points is generated  by a reasonably
well--behaved  point  process  (see  {\pcite{weil:stereology}}  for  a
rigorous  discussion  of  requirements),  we can  define  an  ensemble
average of the  Minkowski functionals per unit volume.   We call these
quantities $v_\mu$.

Like other  morphological estimators,  the Minkowski functionals  of a
point sample depend  on its number density.  In the  case of a Poisson
process one may  derive scaling relations with the  number density and
radius using  the analytical  results of {\cite{mecke:euler}}  and the
homogeneity  property  of  the  Minkowski  functionals.   For  general
distributions,  however, scaling relations  are not  readily available
since the scaling properties depend, even without correlations, on the
dimensionality of  the support of the  point process.  This  is why we
always take care to select the  same number of points from our samples
before  we calculate  the  Minkowski functionals.   However the  total
number of halos above a certain minimum circular velocity changes with
redshift due to merging.  Therefore,  we proceed as follows: We assume
a threshold of $111\kms$  for the circular velocity $v_{\text{circ}}$,
and  find  5575,   9644,  5869  halos  at  redshifts   4,  1,  and  0,
respectively.  From these halos we randomly select at each moment 5000
halos, and  compute their Minkowski  functionals.  We do  this several
times with different random sets  to derive a scatter.  To compare the
distribution of  halos with  that of dark  matter we also  select 5000
dark matter particles at random.

All  the point  samples we  study live  in cubic  boxes  with periodic
boundary conditions.   Hence correction of edge  effects as performed,
for example, by {\scite{schmalzing:minkowski}} in the analysis of real
catalogues, is  not an issue  here.  The numerical computation  of the
Minkowski  functionals is  performed  using an  implementation of  the
partition formula derived by {\scite{federer:curvature}}.  The formula
provides a  generalized concept  of surface integration,  suitable for
taking into account the corners and edges encountered in our union set
of  balls (for  a detailed  description,  see {\pcite{mecke:robust}}).
The present  implementation offers the advantages  of scaling linearly
with the  number of points, and requiring  only moderate computational
resources; a sample of 5000 points  can be fully analyzed in about one
hour on an average workstation.  Together with a parallelized version,
the source of the code used in this work is available from the authors
upon request.

\section{Analysis}
\label{sec:analysis}
\subsection{Evolution of two-point correlation function and Minkowski functionals}

\begin{figure}
\bildchen{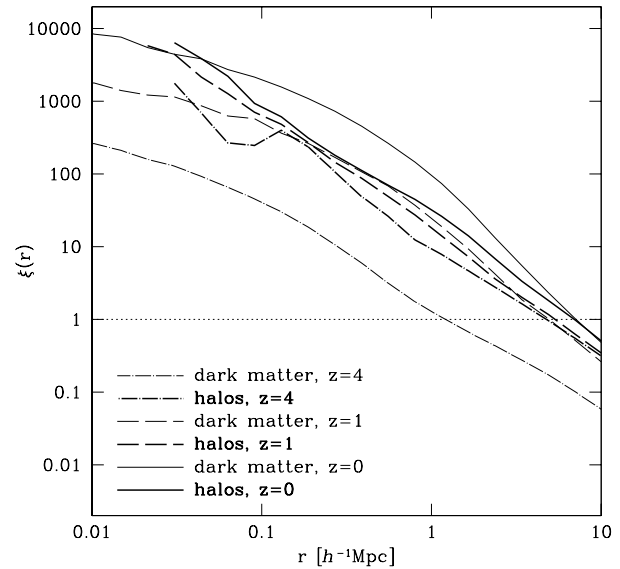}
\caption{
\label{fig:xi}
Correlation functions of the halo samples ($v_{\text{circ}}>111\kms$)
and the dark matter distributions at $z = 0,\; 1,\; 4$.  }
\end{figure}

\begin{figure*}
\bildchen{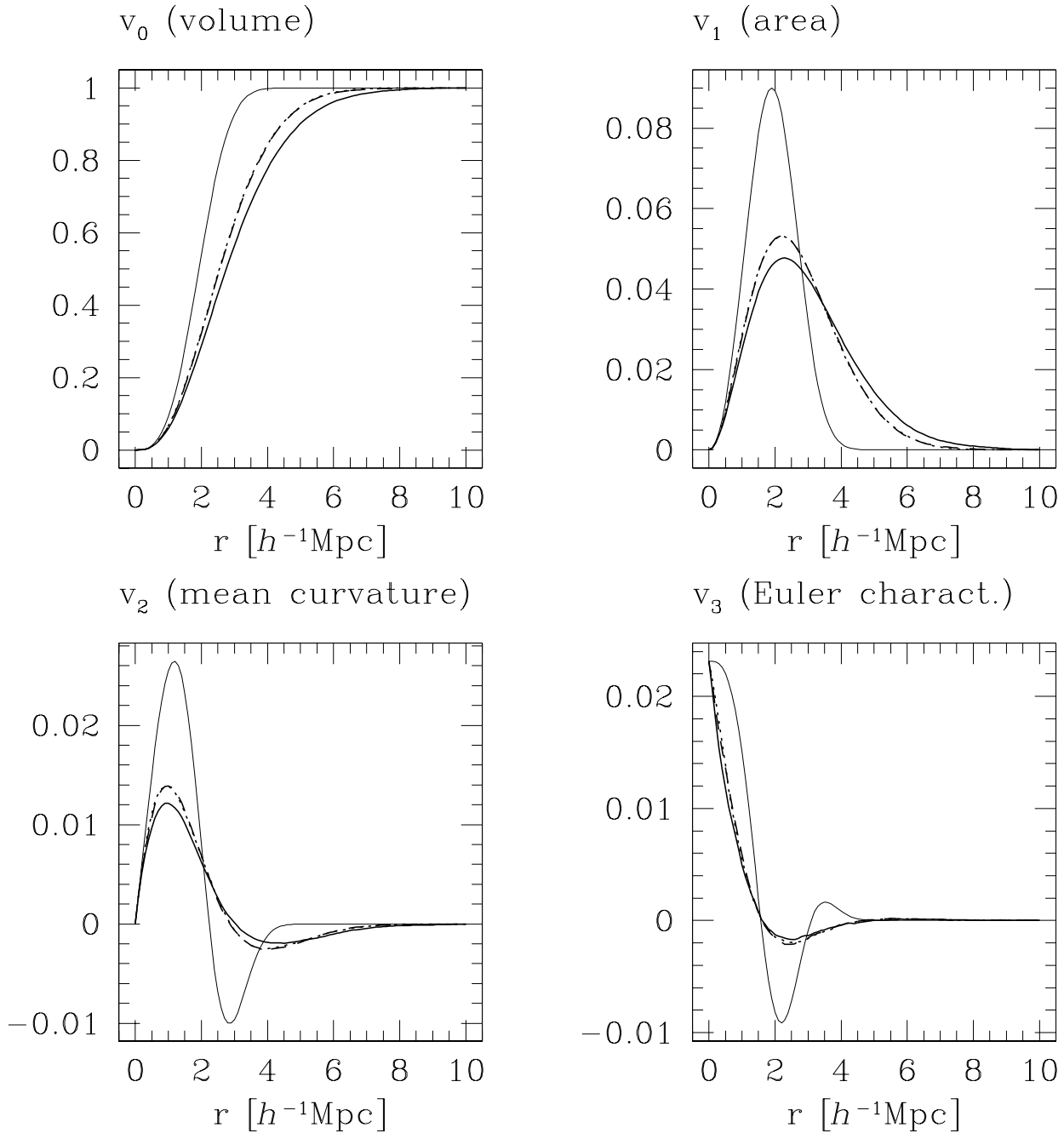}
\caption{
\label{fig:halo}
Minkowski functionals of  the halo samples ($v_{\text{circ}}>111\kms$)
samples  at $z=0$ (solid),  $z=1$ (dashed)  and $z=4$  (dotted).  Note
that the  results for $z=1$  and $z=4$ coincide almost  everywhere, so
the lines  appear dash--dotted.  As  a standard of reference,  we show
the  analytically known  values  for  a Poisson  process  of the  same
intensity (thin solid line). }
\end{figure*}

\begin{figure*}
\bildchen{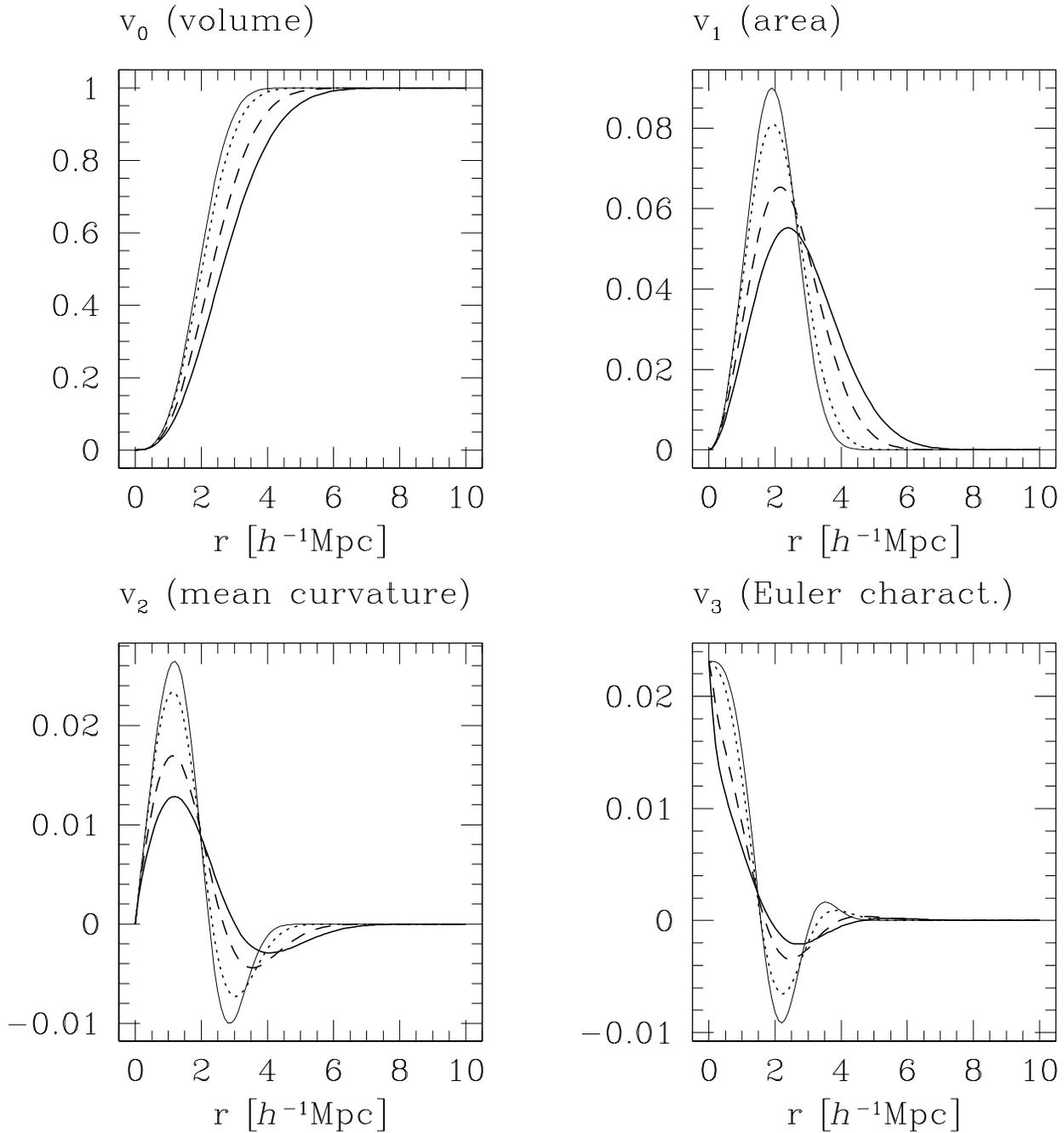}
\caption{
\label{fig:dark}
Minkowski  functionals  of  the  dark  matter  distribution  at  $z=0$
(solid),    $z=1$    (dashed)    and    $z=4$   (dotted).     As    in
Figure~\protect\ref{fig:halo}, we include a Poisson process(thin solid
line).  }
\end{figure*}

Recently, the  evolution of clustering  of dark matter halos  has been
investigated   for  several   cosmological   models  using   numerical
dissipationless                                             simulations
{\cite{brainerd:peculiar,bagla:evolution,colin:evolution,kravtsov:origin,ma:redshift}},
simulations  that   include  both  dissipationless   dark  matter  and
dissipative                     baryonic                    components
{\cite{katz:clustering,blanton:physical,blanton:time,cen:physical,jenkins:evolution}},
and  ``hybrid''  studies  in  which  dissipationless  simulations  are
complemented  with   a  semi-analytical  model   of  galaxy  formation
{\cite{roukema:merging,kauffmann:clusteringI,kauffmann:clusteringII,diaferio:clusteringIII,baugh:modelling,benson:nature,kolatt:}}.
All  of   these  studies,  though  very  diverse   in  their  methods,
qualitatively agree  on one important  result: bias is expected  to be
non--linear,  to depend  on the  properties of  the DM  halos  and the
galaxies they host, and to be a (generally non--monotonic) function of
cosmic time.

In particular, {\scite{colin:evolution}}  discuss the evolution of the
real  space two-point correlation  function in  different cosmological
models.   They  find  that  in  all  models and  at  all  epochs,  the
correlation  function of  halos can  be reasonably  approximated  by a
power--law.   In   Figure~\ref{fig:xi}  we  compare   the  correlation
function  of  the  dark  matter  particles  in  real  space  with  the
correlation    function   of    halos    with   circular    velocities
$v_{\text{circ}}>111\kms$.  In agreement with previous studies, we see
almost no evolution of the halo correlation function between redshifts
4 and  0, whereas by comparison  the correlation function  of the dark
matter particles increases with time.
   
The four Minkowski functionals for  halos and dark matter particles at
various epochs are shown in Figures~\ref{fig:halo} and~\ref{fig:dark},
respectively.  The scatter  between different subsamples is marginally
larger  than the  line width,  so for  clarity we  only plot  the mean
values.  One should keep in mind that due to the density dependence of
the Minkowski  functionals the diagnostic  parameter ``radius'' cannot
be directly  associated with a  scale.  Instead, the  general features
described below occur  within the radius range between  zero and a few
times the average distance for  any point distribution studied.  It is
the location and relative strength of these features that accounts for
the discriminative  power of the  Minkowski functionals.  Quantitative
measurements  may  be  possible  using  reference  distributions;  the
simplest such case is the Poisson process displayed in all Figures.

The  zeroth Minkowski  functional is  nothing  but one  less the  void
probability  function (VPF), which  is defined  as the  probability of
finding no objects within randomly placed balls (for a recent study of
the VPF of halos,  see {\pcite{ghigna:void}}).  Voids are successively
filled  as the  radius  of  balls increases,  so  the $v_0$  increases
monotonically approaching  a value of  one at complete filling  of the
whole simulation box.  The  first and second Minkowski functionals are
related to  the surface content  and the integrated mean  curvature of
the body, and have rarely  been applied in cosmology.  Both quantities
start from a  value of zero at zero radius, and  increase to a maximum
value.  While  the area only decreases  from that point,  as the union
set  becomes more  connected, the  integrated mean  curvature  shows a
turnover from  positive to negative  values, indicating a  change from
convex  to concave  structures.  The  third Minkowski  functional, the
Euler  characteristic  $v_3$, is  a  purely  topological quantity  and
counts  the number  of  components less  the  number of  tunnels in  a
structure.  At small radii, all balls are isolated from each other and
$v_3$ is simply equal to the number density of points.  As connections
are  established   between  closely  neighboring   points,  the  Euler
characteristic  decreases and  eventually  falls below  zero, when  an
intricate network  of tunnels has formed  in the union  set.  A second
maximum is reached at an already fairly large filling factor, when the
union set mainly contains isolated cavities, which count as components
and each contribute $+1$  to the Euler characteristic.  Eventually the
whole  box is  filled and  $v_3$ becomes  zero.  Note  that  the Euler
characteristic  $v_3$ is connected  to the  genus $g$  via $g=1-2v_3$.
The  genus  has been  already  widely  discussed  in the  cosmological
context using density fields constructed by fixed--width smoothing the
particle distribution  {\cite{gott:sponge}}.  The result  then depends
on both the  smoothing scale and the density  threshold which has been
used for the construction of the body.

In  Figure~\ref{fig:halo} we  show the  Minkowski functionals  of 5000
randomly selected halos at redshifts $z=4$,  1, and 0.  We do not show
the  scatter resulting  from different  selection, since  the standard
deviation is of the order of the line width.  Only little evolution of
the clustering  of halos  can be seen.   In particular,  the Minkowski
functionals at  $z=4$ and  $z=1$ coincide, while  the number  of halos
with  $v_{\text{circ}}>111\kms$  increases in  this  time interval  by
almost a factor  of two.  If one calculates  the Minkowski functionals
of all halos  above that threshold of the  circular velocity one would
see weaker clustering  at $z=1$ than at $z=4$.   This demonstrates the
importance of selecting the same  number of objects for the comparison
of Minkowski functionals.

Already  at $z=4$  the  Minkowski functionals  of  halos are  markedly
different from a  Poisson process.  We interpret the  evolution of the
Minkowski  functionals in  the long  time interval  between  $z=1$ and
$z=0$    as    nonlinear    clustering    on   small    scales    (see
{\pcite{kravtsov:origin}} for a  detailed discussion).  The halos tend
to move towards existing clusters of halos, hence the void probability
function increases on  all scales, and both surface  and mean integral
curvature  decrease on  small scales.

It  is   remarkable  that  the   Euler  characteristic  of   the  halo
distribution hardly  changes with time.   The absence of  the positive
maximum of $v_3$  at any time and any radius  in the halo distribution
can  be interpreted  as the  signature of  a network  of  filaments or
sheets, which leave only few  cavities in the almost filled union set.
Furthermore, the sharp drop of the Euler characteristic at small radii
points  towards strong  clustering  on small  scales.  These  features
reveal the  presence of a  network of filamentary  structures spanning
the whole  simulation box.  We are  confident that our  result is only
weakly  influenced by  the comparatively  small box  size  of 60\hMpc.
After  all, two  points in  the set  can only  influence  each other's
contribution to the Minkowski functionals  if they are less than twice
the  radius apart.   However, complete  filling occurs  below 10\hMpc,
i.e. almost an order of magnitude below the box size.

In contrast to the modest evolution seen in the halo distribution, the
clustering  of  dark  matter  shown in  Figure~\ref{fig:dark}  evolves
considerably with time.   The present day clustering of  DM appears to
be as strong as that of the $z=0$ halo sample.  At $z=4$, however, the
DM clustering properties still  resemble those of the Poisson process.
The modest differences from  a purely random distribution reflect that
most of  the particles are  still contained in  the field, and  only a
minority  has formed  structures.   Even if  the  sampling density  is
increased, this effect persists.

Finally, let us  compare the Minkowski functionals of  the DM and halo
distributions.   The  DM  distribution  evolves  continuously  towards
higher clustering  as one  has seen already  from the analysis  of the
evolution  of the  correlation function  and the  power  spectrum.  At
early  epochs  the  Minkowski  functionals indicate  a  much  stronger
clustering  of the  halos  than of  DM,  but there  is  only a  little
difference in the morphological properties of halo and DM distribution
at $z=0$ (cf. Sect.~\ref{sec:bias}).

\subsection{Comparison to Correlation Function Analysis}

\begin{figure}
\bildchen{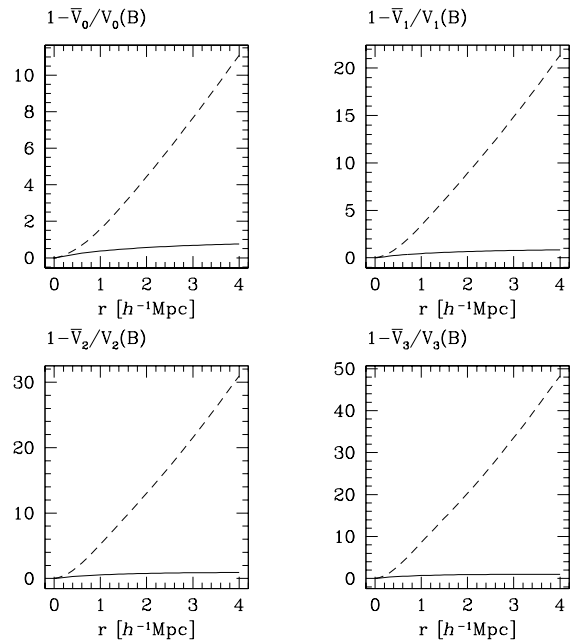}
\caption{
\label{fig:xi-mini}
For   halos  at   $z=0$  we   compare  the   asymptotic   series  from
Eq.~(\protect\ref{eq:hierarchyseries}) (solid  line) to its truncation
after the first non--trivial term, which contains a convolution of the
two point correlation function.  The dashed line is obtained using the
halo correlation function of Fig.~(\ref{fig:xi}).  }
\end{figure}

\begin{table}
\begin{equation*}
\begin{alignat*}{3}
&\mu&\qquad&V_\mu(B)&\qquad&\frac{V_\mu\left(B\cap{}B_\bx\right)}{V_\mu(B)}\\
\\ \hline \\
&0  && 4\pi r^3/3 && (1-d)^2(1-d/2) \\
&1  && 2\pi r^2/3 && 1-d \\
&2  && 4r/3         && 1-d+ (\sqrt{1-d^2}/2)\arcsin{d} \\
&3  && 1 && 1 
\end{alignat*}
\end{equation*}
\caption{
\label{tab:intersect}
Minkowski functionals of a single ball $B$ of radius $r$ and of an
intersection $B\cap{}B_\bx$ of two balls separated by a distance
$x=2rd$ ($0\le d \le 1$).}
\end{table}

In theory, Minkowski functionals contain information about correlation
functions of arbitrary order.  However,  it is not clear a priori that
this actually results in enhanced information compared to the standard
correlation function analysis.  Although  the analysis of the previous
section  revealed  a  wealth   of  information  on  the  evolution  of
clustering  in our simulation,  it seems  that qualitatively  the same
information  could have  been  deduced from  the correlation  function
alone.   In Appendix~\ref{sec:hierarchy}  we  establish an  analytical
connection  between the  hierarchy  of correlation  functions and  the
Minkowski  functionals, and  use  this relation  to  conduct a  direct
comparison.

The  Minkowski functionals  per  unit volume,  $v_\mu$,  of a  Poisson
process with intensity $\rho$ are given by
\begin{equation}
\begin{split}
v_0 &= 1-\re^{-\rho{V}_0(B)} \\
v_1 &= \re^{-\rho{V}_0(B)} \rho{V}_1(B) \\
v_2 &= \re^{-\rho{V}_0(B)} \left( \rho{V}_2(B)-\frac{3\pi}{8}\rho^2{V}_1(B)^2 \right) \\
v_3 &= \re^{-\rho{V}_0(B)} \left( \rho{V}_3(B)-\frac{9}{2}\rho^2{V}_1(B){V}_2(B)+\frac{9\pi}{16}\rho^3{V}_1(B)^3 \right)
\end{split}
\label{eq:poisson}
\end{equation}
where $V_\mu(B)$ is the $\mu$th  Minkowski functional of a ball $B$ of
radius $r$.  As shown in Appendix~\ref{sec:hierarchy} this formula can
be generalized  for arbitrary point  processes, if the  $V_\mu(B)$ are
replaced by  a series $\lV_\mu$  containing the complete  hierarchy of
correlation functions.  Truncating Equation~(\ref{eq:hierarchyseries})
after the first two terms we have
\begin{equation}
\lV_\mu=V_\mu(B)-\frac{\rho}{2}\int\rd^3x\xi(x)V_\mu(B\cap{B_x})\pm\ldots
\label{eq:truncatedseries}
\end{equation}

In  Table~\ref{tab:intersect} we  summarize the  Minkowski functionals
$V_\mu(B)$ of a  single ball $B$ of radius $r$  and of an intersection
$B\cap{}B_\bx$ of two  balls separated by a distance  $x=2rd$ ($0\le d
\le  1$).    Note,  that  all  Minkowski  functionals   of  the  empty
intersection $x>2r$ become zero.

In Figure~\ref{fig:xi-mini} we present the asymptotic series $\lV_\mu$
of the halo  distribution at $z=0$ (solid line).   We compare the full
asymptotic         series         obtained        by         inverting
Eq.(\protect\ref{eq:minfrombare}) with its  truncation after the first
non--trivial      term       (dashed      line)      according      to
Eq.~(\protect\ref{eq:truncatedseries}).    Here   we   use  the   halo
two-point correlation function at $z=0$ determined from the simulation
as presented  in Fig.~\ref{fig:xi}.   Obviously, the leading  terms do
not    even    vaguely    approximate    the    full    series    from
Equation~(\ref{eq:truncatedseries}).

Integrating  Eq.~(\ref{eq:truncatedseries}) with a  standard power-law
correlation  function  $\xi(x)=  x^{-\gamma}$  we obtain  the  scaling
behavior
\begin{equation}
1-\lV_\mu/V_\mu(B)\propto r^{3-\gamma}
\end{equation}
for the  truncated series.   For clarity, we  have not  included these
curves in Fig.~\ref{fig:xi-mini},  but with $\gamma=1.7$, a reasonable
fit  to   the  curve   calculated  from  the   numerically  determined
correlation  function  can  be   obtained.   This  confirms  that  the
Minkowski  functionals  reveal   information  beyond  the  correlation
function   analysis,  and  sharply   discriminate  between   the  halo
distribution of our model and  a purely Gaussian process with the same
second--order   characteristics.    Since   they   reveal   clustering
properties of  a totally different quality,  Minkowski functionals are
an ideal complement to more standard methods of large--scale structure
analysis.

\subsection{Bias}
\label{sec:bias}

\begin{figure}
\bildchen{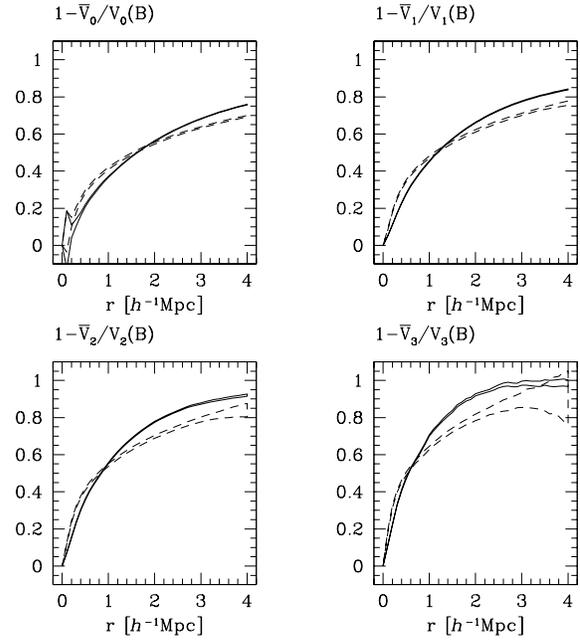}
\caption{
\label{fig:bias.z0}
In these plots we compare the ratios $1-\lV_\mu/V_\mu(B)$ for halos
(solid lines) and dark matter particles (dashed lines) at redshift
$z=0$.  The areas indicate the standard deviation between different
subsamples of the point distribution.  }
\end{figure}

\begin{figure}
\bildchen{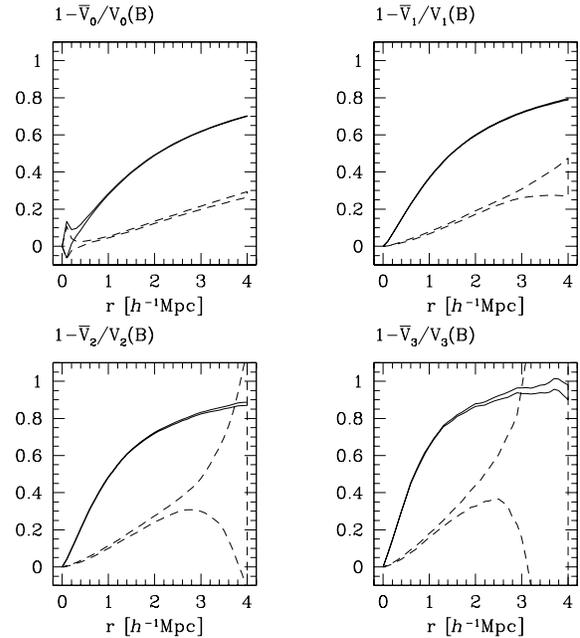}
\caption{
\label{fig:bias.z4}
As in Figure~\protect\ref{fig:bias.z0}, these plots show the ratios
$1-\lV_\mu/V_\mu(B)$ for halos (solid lines) and dark matter particles
(dashed lines), but at redshift $z=4$.  Again, the areas display the
scatter between subsamples of 5000 points taken from the whole
particle distribution.  }
\label{lastpage}
\end{figure}

The  comparison  of  Minkowski  functionals  on  its  own  provides  a
qualitative understanding  of the  clustering properties of  matter in
the  model.   In  this  section  we compare  directly  the  clustering
properties  of DM  and  halos at  $z=0$.  It is  well  known that  the
distribution  of halos  is  different from  the  DM distribution.  The
relationship    between   these    two    distributions   is    called
bias. Considering the correlation functions  of halos and DM the ratio
between   both   can   be   interpreted   as   scale--dependent   bias
(Fig.~\ref{fig:xi}).

This  scale--dependent bias  is a  generic prediction  of hierarchical
models of  structure formation {\cite{colin:evolution}}.   The bias is
also  time-dependent: it  decreases from  an initially  high  value at
$z\sim3-7$ to  a considerably  smaller value (which  can be  even less
than one) at present.  Qualitatively, the same behavior has been found
in the power spectrum  both in real space {\cite{kravtsov:origin}} and
in redshift space  {\cite{gottloeber:halo}}.  Since power spectrum and
correlation  function are the  Fourier pairs,  a bias  in one  of them
leads to a bias in the other.  Obviously, any scale dependence will be
different in coordinate and $k$ space.

In  Figures~\ref{fig:bias.z0} and  {\ref{fig:bias.z4}} we  compare the
functions  $1-\lV_\mu/V_\mu(B)$ for  the dark  matter and  halos.  Our
analysis  indicates   that  these   functions  (the  solid   lines  in
Figure~\ref{fig:xi-mini}) are  much more suitable  for comparison than
the     Minkowski     functionals     themselves.     For     example,
$1-\lV_\mu/V_\mu(B)=0$ is expected in  case of a Poisson process.  The
figure shows these quantities for  halos (solid lines) and dark matter
particles  (dashed lines).  At  $z=0$, bias  is a  lot weaker  than at
$z=4$.  For small radii, an anti--bias is observed, meaning that halos
appear less clustered than dark matter particles.  This does not occur
at  higher redshifts, where  halos cluster  stronger than  dark matter
particles at all radii.  Although the  radii for which the halo and DM
curves coincide  are different for the four  functionals and different
from  the radius  where  the  correlation functions  of  halos and  DM
coincide, the  qualitative appearance of all  Minkowski functionals is
similar;  this reflects  the fact  that  all $\lV_\mu$  depend on  the
hierarchy  of correlation  functions  in a  very  similar way  through
Equation~(\ref{eq:hierarchyseries}).

The two  solid and  dashed lines  in each of  the panels  indicate the
scatter  which we  have estimated  from the  statistics  for different
random  selections of  5000  objects.  At  large radii,  uncertainties
increase because the Boolean  grain model already fills a considerable
portion of the whole space, and small fluctuations in the pattern have
strong  effects on  its morphology.   This  effect is  expected to  be
strongest for  structures that eventually  form cavities in  an almost
solid  body.  In  fact the  dark matter  distribution shows  much more
scatter  at larger  radii,  compared to  the  fairly filamentary  halo
distribution.   While the  scatter described  so far  is  an intrinsic
feature  of the point  process, the  very large  error for  the volume
functionals at small radii is  due to the Monte Carlo integration used
for the numerical evaluation; by  increasing the number of Monte Carlo
points, it can be somewhat reduced but not fully eliminated.

It  is  interesting to  discuss  the  anti-bias  of halo  distribution
observed in all  of the Minkowski functionals at  $z=0$. Note that our
halo finder algorithm identifies both isolated halos and halos located
within    virial    radii   of    other    halos   (satellites;    see
{\pcite{colin:evolution}}  for details).  Therefore,  the small--scale
anti-bias is  not caused by the  characteristic size of  the halos and
their mutual exclusion. Note, for example, that there is no indication
of   similar  anti-bias   at   $z=4$.   Rather,   as   was  shown   by
{\scite{kravtsov:origin}},  the  anti--bias  is  most  likely  due  to
dynamical friction and (to a  lesser degree) to the tidal stripping in
the high--density  environments of  groups and clusters.   Indeed, the
dynamical  friction results  in mergers  of satellite  halos  with the
central cluster halo and thereby reduces the ratio of halo overdensity
to the ever--growing  overdensity of dark matter. We  refer the reader
to {\scite{kravtsov:origin}}  for an extended discussion  and tests of
the origin and evolution of halo  anti-bias.  Here we note that if the
anti--bias is  indeed due to the dynamical  processes affecting galaxy
evolution   in   groups  and   clusters,   studies  of   higher--order
small--scale galaxy  clustering and  its dependence on  morphology may
prove to be useful and independent probes of galaxy evolution in dense
environments.

\section{Summary and conclusions}
\label{sec:conclusions}

We  have used  a  high--resolution dissipationless  simulation of  the
formation of dark matter halos  in a flat cosmological model dominated
by the cosmological constant and cold dark matter.  In the $60h^{-1}{\
}{\rm Mpc}$  simulation box  we have identified  5575, 9644,  and 5869
halos  of  maximum   circular  velocity  $v_{\text{circ}}>111\kms$  at
redshifts 4, 1, and 0, respectively.

At each of  these epochs we have calculated  the Minkowski functionals
for  bodies constructed  as a  Boolean  grain model  of 5000  randomly
selected  halos  and 5000  randomly  selected  dark matter  particles,
respectively.  The  Minkowski functionals focus on  global features of
three--dimensional  bodies,  and  provide  a unifying  frame  for  the
analysis  of cosmic  structures which  comprises the  void probability
function as  well as the  genus statistics.  Minkowski  functionals do
not replace  but complement such  traditional tools as  the two--point
correlation  function,  since  they  reveal clustering  properties  of
higher-order.

We found  almost no  evolution of the  Minkowski functionals  of halos
between $z=4$ and $z=1$ and  little evolution between $z=1$ and $z=0$.
The  rather small  differences at  various  epochs are  mainly due  to
evolution of  nonlinear clustering on small scales.   Already at $z=4$
the  Minkowski functionals  of  halos are  markedly  different from  a
Poisson  process.  We  can deduce  that from  the  beginning, luminous
matter quickly forms complicated  structure on large scales, which may
be interpreted as a network of sheets and filaments.

On the other hand, the  clustering of dark matter evolves considerably
with time.  At  $z=4$ the clustering properties of  DM mostly resemble
those  of the  Poisson process,  with only  minor deviations.   At the
present epoch,  the clustering  is as strong  as for the  halo sample,
although the  morphology of the dark matter  distribution differs from
the structure formed by the halos.

The difference of  the Minkowski functionals of halos  and dark matter
can be  interpreted as a  scale dependent bias which  is qualitatively
similar  but  quantitatively  different  for  each  of  the  Minkowski
functionals.   The evolution  of halo  bias in  the linear  and mildly
nonlinear regimes results from an interplay between halo formation and
merging rates in  different regions and in the  field.  Both rates are
non-monotonic functions of time, but their combined effect is decrease
of the  initially high {\em  statistical\/} bias during the  course of
the evolution.  In the non-linear  regime, the halo bias evolution may
be also affected by the dynamical processes inside clusters and groups
{\cite{kravtsov:origin}}.

We  have demonstrated  in this  paper that  the  Minkowski functionals
contain  information  about  the  correlation functions  of  arbitrary
order. Therefore, our results imply that the scale-dependent bias is a
property of not only the two-point halo correlation function, but also
of the  correlation functions of  higher order.  This, in  turn, means
that  there  are morphological  differences  between distributions  of
halos and  matter at  both small and  large scales. In  particular, at
early  epochs  the halos  are  already  strongly  clustered and  their
distribution shows  distinct features of  a network.  The  dark matter
distribution at the same epochs is still weakly clustered and is close
to  the   Poisson  distribution.   At  the  present   epoch  the  halo
distribution appears less clustered  on small scales, while it remains
more  filamentary on  large scales  than the  distribution  of matter.
Whether the  measured Minkowski functionals indicate  the abundance of
filaments  or sheets  or  both in  the  network is  currently an  open
question.   Although the  very steep  initial  rise of  the curves  in
Figure~\ref{fig:bias.z0}  would   favor  a  filamentary  distribution,
sheet--like  structures  are   not  ruled  out  either.   Quantitative
measurements of planarity  and filamentarity will be the  subject of a
follow--up article.

\section*{Acknowledgments}

This work  was funded  by grants from  the NSF  and NASA to  NMSU.  We
acknowledge  support by the  NATO grant  CRG 972148.   SG acknowledges
support from  the Deutsche Akademie der  Naturforscher Leopoldina with
means of  the Bundesministerium f\"ur Bildung und  Forschung grant LPD
1996.  JS wishes to thank the  AIP for hospitality during a stay where
parts of  this work were  prepared. The simulation presented  here was
performed at the National Center for Supercomputer Applications (NCSA,
Urbana-Champaign,  Illinois) and  on  the Origin2000  computer at  the
Naval Research Laboratory.

Please note that the code used for this analysis is publicly available
and    can     be    obtained    by    sending     email    to    {\tt
jensen@mpa-garching.mpg.de}.

% Bibliography generated by bibtex
\providecommand{\bysame}{\leavevmode\hbox to3em{\hrulefill}\thinspace}

\appendix

\section{Minkowski functionals expressed in terms of correlation functions}
\label{sec:hierarchy}

It is  known that Minkowski functionals  contain correlation functions
of  arbitrary order.   Here we  aim  to establish  a precise  relation
between the  average Minkowski functionals $v_\mu$ per  unit volume in
terms of the  hierarchy of correlation functions $\xi_n$  of the point
process.

In order to cut down on notation we always use a linear combination of
Minkowski  functionals,  the  so--called Minkowski  polynomial  $M(t)$
{\cite{hadwiger:wills,kellerer:minkowski}}.  We define
\begin{equation}
M(t;K)=\sum_{\mu=0}^d\frac{\alpha_{\mu}t^\mu}{\mu!}V_\mu(K)
\end{equation}
with  coefficients\footnote{$\omega_\mu$  denotes  the volume  of  the
$\mu$--dimensional  unit  ball.   Some  important special  values  are
$\omega_0=1$,         $\omega_1=2$,         $\omega_2=\pi$,        and
$\omega_3=\frac{4}{3}\pi$.}
\begin{equation}
\alpha_\mu=\frac{\omega_{d-\mu}}{\omega_d}.
\end{equation}
Analogously,  we  construct a  polynomial  $m(t)$  from the  densities
$v_\mu$ of the Minkowski functionals per unit volume.  From now on, we
drop the  variable $t$  from the polynomial,  and only  reintroduce it
when extracting the individual Minkowski functionals via
\begin{equation}
V_\mu(K)=\frac{1}{\alpha_{\mu}}\frac{\partial^\mu}{\partial{t}^\mu}M(t;K).
\label{eq:minfrompolynomial}
\end{equation}

By taking  the ensemble average  and applying the  additivity relation
repeatedly, we obtain the formula given by {\scite{mecke:robust}},
\begin{multline}
m=1-
\sum_{n=0}^\infty\frac{(-1)^n}{n!} \\
\int\rd\tau_n
\rho_n\left(\bx_1,\ldots,\bx_n\right)
M\left(B_{\bx_1}\cap\ldots\cap{}B_{\bx_n}\right),
\label{eq:minfromrho}
\end{multline}
with the integration measure defined by
\begin{equation}
\int\rd\tau_n=\int_\cD\rd^dx_1\ldots\rd^dx_n.
\end{equation}

If the  density functions $\rho_n$  were independent of  position, the
integrals could be performed  using the principal kinematic formula by
{\scite{blaschke:I}} (see also {\pcite{chern:cinematica}} for the more
general version  employed here).   Fixing one body  $A$ in  space, and
intersecting with a moving body  $B$, we obtain after integration over
all possible positions of $B$ a simple factorization.
\begin{equation}
\int_\cD\ddx M(A\cap{B_\bx})=M(A)M(B),
\end{equation}
where equality  holds to order  $t^d$.  Unfortunately, straightforward
application   of   the  principal   kinematic   formula  to   Equation
(\ref{eq:minfromrho})  is  impossible  in  most situations  since  the
density functions  are in general position dependent.   However, it is
possible to use the  factorization properties of the density functions
into the correlation functions.

Equation  (\ref{eq:minfromrho})   motivates  the  introduction   of  a
hierarchy  of ``moments''  $P_n$ by  weighting the  $n$--point density
function with the Minkowski polynomial of $n$ intersecting balls,
\begin{equation}
P_n
=
\int\rd\tau_n
\rho_n\left(\bx_1,\ldots,\bx_n\right)
M\left(B_{\bx_1}\cap\ldots\cap{}B_{\bx_n}\right).
\end{equation}
Writing  the density  functions  $\rho_n$ as  sums  over all  possible
partitions into correlation functions $\xi_n$, and using the principal
kinematic   formula,  it   can   be  shown   that  the   ``cumulants''
corresponding to the moments $P_n$ are the $\Xi_n$ given by
\begin{equation}
\Xi_n
=
\int\rd\tau_n
\xi_n\left(\bx_1,\ldots,\bx_n\right)
M\left(B_{\bx_1}\cap\ldots\cap{}B_{\bx_n}\right).
\end{equation}

Given  a generating  functional  $P[j]$ for  the  $P_n$, a  generating
functional $\Xi[j]$ for the $\Xi_n$ is obtained through
\begin{equation}
P[j]=\exp\left(\Xi[\rho j]\right).
\end{equation}
Setting $j=-1$, we directly arrive at the result
\begin{equation}
m=
1-\sum_{n=0}^\infty\frac{(-1)^n}{n!}P_n=
1-\exp\left(\sum_{n=0}^\infty\frac{(-\rho)^n}{n!}\Xi_n\right).
\end{equation}

Recovering the Minkowski functionals  from the polynomial via Equation
(\ref{eq:minfrompolynomial}), the relation in $d=3$ dimensions reads
\begin{equation}
\def\expo{\exp\left(-\rho\lV_0\right)}
\begin{split}
v_0 &= 1-\expo \\
v_1 &= \expo\rho\lV_1 \\
v_2 &= \expo\left(\rho\lV_2-\frac{3\pi}{8}\rho^2\lV_1^2\right) \\
v_3 &= \expo\left(\rho\lV_3-\frac{9}{2}\rho^2\lV_1\lV_2+
\frac{9\pi}{16}\rho^3\lV_1^3\right)
\label{eq:minfrombare}
\end{split}
\end{equation}
which looks remarkably similar to  the result for the Poisson process,
except  that the bare  Minkowski functionals  of balls  $V_\mu(B)$ are
replaced by the quantities
\begin{multline}
\lV_\mu=V_\mu(B)+
\sum_{n=1}^\infty
\frac{(-\rho)^n}{(n+1)!}
\int_\cD\ddx_1\ldots\ddx_n \\
\xi_{n+1}(0,\bx_1,\ldots,\bx_n)
V_\mu(B\cap{B_{\bx_1}}\cap\ldots\cap{B_{\bx_n}}).
\label{eq:hierarchyseries}
\end{multline}
Note that  Equation~(\ref{eq:hierarchyseries}) is a  generalization of
the  result  by  {\scite{white:hierarchy}}  for the  void  probability
function.   This shows that  all Minkowski  functionals depend  on the
complete  hierarchy  of correlation  functions  through an  asymptotic
series,  hence   each  order  $n$   can  be  expected   to  contribute
significantly  to the  overall result.   In Section~\ref{sec:analysis}
this claim is backed by a direct comparison of numerical results.

While  it  looks promising  to  push  these analytical  considerations
forward towards a  hierarchical model {\cite{balian:I,balian:II}} or a
lognormal  Cox process  {\cite{coles:lognormal}}, we  will  stop here,
since the series to second order is sufficient for our purposes.

\label{lastpage}

\end{document}